\documentclass[aps,prb,showpacs,twocolumn,floats]{revtex4}
\usepackage{amssymb}
\usepackage{amsbsy}
\usepackage{amsmath}
\usepackage{graphicx}
\usepackage{amsmath}
\usepackage{times}
\usepackage{color}
\usepackage{subfigure}
\usepackage{setspace}
\usepackage{bm}
\newcommand\bea{\begin{eqnarray}}
\newcommand\eea{\end{eqnarray}}
\newcommand\beq{\begin{equation}}
\newcommand\eeq{\end{equation}}
\newcommand\bib{\bibitem}

\begin{document}

\title{Tuning towards dynamic freezing using a two-rate protocol}

\author{Satyaki Kar, Bhaskar Mukherjee, and K. Sengupta}

\affiliation{Theoretical Physics Department, Indian Association for
the Cultivation of Science, Jadavpur, Kolkata-700032, India.}

\date{\today}

\begin{abstract}

We study periodically driven closed quantum systems where two
parameters of the system Hamiltonian are driven with frequencies
$\omega_1$ and $\omega_2=r \omega_1$. We show that such drives may
be used to tune towards dynamics induced freezing where the
wavefunction of the state of the system after a drive cycle at time
$T= 2\pi/\omega_1$ has almost perfect overlap with the initial
state. We locate regions in the $(\omega_1 ,r)$ plane where the
freezing is near exact for a class of integrable and a specific
non-integrable model. The integrable models that we study encompass
Ising and XY models in $d=1$, Kitaev model in $d=2$, and Dirac
fermions in graphene and atop a topological insulator surface
whereas the non-integrable model studied involves the experimentally
realized one-dimensional (1D) tilted Bose-Hubbard model in an
optical lattice. In addition, we compute the relevant correlation
functions of such driven systems and describe their characteristics
in the region of $(\omega_1,r)$ plane where the freezing is
near-exact. We supplement our numerical analysis with semi-analytic
results for integrable driven systems within adiabatic-impulse
approximation and discuss experiments which may test our theory.

\end{abstract}


\maketitle

\section{Introduction}
\label{intro}

The behavior of closed quantum systems driven out of equilibrium has
attracted a lot of attention in recent times
\cite{pol1,dziar1,dutta1}. Such out-of-equilibrium systems explore a
major region (if not all) of it's Hilbert space leading to a gamut
of novel phenomena which have no equilibrium counterparts. For
example, they exhibit universal scaling of excitation density when
driven slowly through a quantum critical point
\cite{kz1,pol2,ks1,ks2,pol3, sondhi1}. Another such class of
phenomenon involves dynamic transition where the free energy density
of a quantum system after a quench develops non-analyticities
leading to sharp cusp-like features in it's Lochsmidt echo pattern
at specific times \cite{mukh1,pol4,dytr1,dytr2,dutta2}; these
transitions have no equilibrium analogs and can not be described by
any local order parameter. Furthermore, the universality and the
possibility of application of renormalization group (RG) methods to
such driven systems has been a subject of recent research
\cite{ehud1, sang1,aditi1}; such analysis also leads to a several
interesting features involving RG flow and universality of driven
systems which are qualitatively different from their equilibrium
counterparts.

A class of such dynamical phenomenon involve periodically driven
quantum systems leading to multiple passages through their critical
points during a drive cycle \cite{stu1,dyn1,dyn2,adas1,rev1}. Such
systems exhibit a wide class of dynamical phenomena which do not
occur in their aperiodically driven counterparts. For example, they
display signatures of Stuckelberg interference phenomenon in their
excitations densities,  dynamical correlation functions, and
thermodynamic quantities such as work distribution \cite{dyn2,ks3}.
Furthermore, periodically driven integrable systems show a separate
class of dynamical transitions which originates from the change in
topology of their Floquet spectrum and leaves its imprint on
temporal behavior of local correlation functions \cite{sen1}. Such
driven systems also exhibit the phenomenon of dynamics induced
freezing \cite{adas1,pekker1,uma1,rev1}; this phenomenon manifests
itself in a near unity overlap of the system wavefunction after
single or multiple drive period(s) with the initial wavefunction at
$t=0$. Such near-unity overlap signifying freezing occurs at
specific discrete frequencies $\omega^{\ast}$ and is therefore
qualitatively different from the trivial freezing that occurs for
ultrafast drive frequencies.

In all of the phenomena mentioned above, the drive protocols used
involve quench, ramp, or periodic change of a single parameter of
the system Hamiltonian as a function of time. More recently, a
separate class of protocols involving linear or power law ramp of
two Hamiltonian parameters with distinct rate has been studied in
details \cite{jay1,rev1,uma1}. The novel features of dynamics
induced by such two-rate ramp protocols include a generalization of
the Kibble-Zurek scaling laws and possibility of reduction of
excitation production as the system is ramped through a quantum
critical point. The latter feature leads to better fidelity for
quantum state preparation and provide a controlled way of reduction
of heating during dynamics in critical systems \cite{rev1}. To the
best of our knowledge, the study of dynamics of closed quantum
critical systems for such two rate protocols has been carried out
for ramp protocols only; no such studies have been undertaken for
periodic drives.

In this work, we aim to fill up this gap in the literature by
studying the effect of two-rate periodic drive protocols on both
integrable and non-integrable closed quantum systems. The drive
protocol chosen involves periodic variation of two parameters of the
system Hamiltonian with frequencies $\omega_1$ and $\omega_2 = r
\omega_1$. For integrable models, we choose a class of models
represented by free fermionic Hamiltonian $H=  \sum_{\vec k}
\psi_{\vec k}^{\dagger} H_{\vec k} \psi_{\vec k}$, where $\psi_{\vec
k} = (c_{\vec k}, c_{-\vec k}^{\dagger})^T$ is the two component
fermion field, $c_{\vec k}$ denotes fermionic annihilation operator,
and $H_{\vec k}$ is a $2 \times 2 $ matrix Hamiltonian density given
by
\begin{eqnarray}
H_{\vec k} &=& \left[ \lambda_1(\omega_1 t) - b_{\vec k}\right]
\tau_3 + \lambda_2(r \omega_1 t) \Delta_{\vec k} \tau_1
\label{fermhamden}
\end{eqnarray}
where $\tau_3 $ and $\tau_1$ are Pauli matrices in particle-hole
space. Such free fermionic Hamiltonians represents Ising and XY
models in $d=1$ \cite{subir0} and Kitaev model in $d=2$
\cite{kit1,sen2,feng1}. In addition, it also describes Dirac-like
quasiparticles in graphene \cite{grarev1} and on the surface of a
topological insulators \cite{toprev1}. In what follows we shall
carry out numerical analysis of this model in the context of XY
model in $d=1$; we note, however, that our results shall be valid
for any other representations of $H_{\vec k}$.

The XY model in a transverse field constitute a model for of
half-integer spins on a one-dimensional (1D) chain whose Hamiltonian
is given by
\begin{eqnarray}
H_{\rm XY} = \sum_{\langle i j\rangle, \alpha=x,y} J_{\alpha}
S_i^{\alpha} S_j^{\alpha} - h \sum_i S_i^z,  \label{xyham1}
\end{eqnarray}
where $J_{x,y}$ are nearest coupling between $x$ and the $y$
components of the spins, $\langle ij \rangle$ indicates that $j$ is
one of the neighboring sites of $i$, and $h$ is the transverse
field. A standard Jordan-Wigner transformation \cite{dutta1} maps
Eq.\ \ref{xyham1} to Eq.\ \ref{fermhamden} in $d=1$ with the
identification
\begin{eqnarray}
b_k &=& (J_x + J_y) \cos(k), \quad \lambda_1= -h
\nonumber\\
\Delta_k &=& - i\sin(k), \quad \lambda_2= (J_x-J_y) \label{mapxy}
\end{eqnarray}

For non-integrable system, we choose the recently experimentally
realized ultracold boson in their Mott insulating phase in the
presence of an artificial electric field created either by shifting
the center of the confining harmonic trap or by applying the
spatially varying Zeeman magnetic field to the spinor bosons
\cite{subir1,subir2,ksen1,ksen2,uma1,bakr1}. The Hamiltonian of such
bosons can be written as
\begin{eqnarray}
H_1 &=& -J \sum_{\langle i j\rangle} (b_i^{\dagger} b_j + {\rm h.c})
+ \sum_i \left[ \frac{U}{2} n_i (n_i -1) -E n_i i \right]\nonumber\\
\label{tiltham1}
\end{eqnarray}
where $b_i$ denotes boson annihilation operator on site $i$, $U$ is
the on-site interaction, $E$ is the magnitude of the effective
electric field in units of energy \cite{subir1}, $J$ is hopping
amplitude of bosons between the nearest site, and $n_i=
b_i^{\dagger} b_i$ is the boson number operator. It has been shown
in Ref.\ \onlinecite{subir1} that the low-energy properties of $H_1$
is captured by a dipole model whose Hamiltonian is given by
\begin{eqnarray}
H_d=-J \sqrt{n_0(n_0+1)} \sum_{\ell}
(f_{\ell}^{\dagger}+f_{\ell})+(U-E) \sum_{\ell} n_{\ell},
\label{diham1}
\end{eqnarray}
where $f_{\ell}=b_i^{\dagger} b_j/{\sqrt{n_0(n_0+1)}}$ is the dipole
annihilation operator living on the link $\ell$ between sites $i$
and $j$, and $n_0$ denotes the boson occupation number of the parent
Mott phase. These dipoles constitute bound states of a boson and a
hole on adjacent sites $i$ and $j$ of the lattice and
$n_{\ell}=f_{\ell}^{\dagger} f_{\ell}$ is the dipole number operator
residing on the link $\ell$ between these sites. The dipole
operators satisfy two additional constraints; first, at most one
dipole can occur on any link ($n_{\ell} \le 1$) and second, the
maximal number of dipoles on two adjacent links is also unity
($n_{\ell} n_{\ell \pm 1}=0$). The model has two ground states; for
$U \gg E$, the state correspond to the dipole vacuum which is same
as the parent Mott state of the bosons while for $ U \ll E$, the
state corresponds to maximal number of dipoles which is a density
wave state for the bosons. It was shown in Ref. \onlinecite{subir1}
that these two ground states are separated by a transition at $E_c =
U + 1.31 J \sqrt{n_0(n_0+1)}$ which belongs to the Ising
universality class. These phases of the model and the signature of
the Ising transition between them have been verified in a recent
experiment \cite{bakr1}. The non-equilibrium dynamics of the model
for single parameter quench, ramp and periodic protocols have also
been studied \cite{ksen1,ksen2,uma1}. In what follows, we shall vary
the electric field $E$ and hopping amplitude $J$ of these models
according to periodic protocols characterized by frequencies
$\omega_1$ and $ \omega_2=r \omega_1$ respectively.

The main results that we obtain from such a study are the following.
First, we develop an analytical framework for studying dynamics for
integrable models (Eq.\ \ref{fermhamden}) subjected to arbitrary
two-rate periodic protocol characterized by frequencies $\omega_1$
and $\omega_2=r \omega_1$ using an adiabatic-impulse approximation
\cite{nori1}; the formalism developed constitutes a generalization
of the adiabatic-impulse approximation method to two-rate periodic
protocols. We show that in the low-frequency regime, where this
approximation is expected to be accurate, our results match those
obtained using exact numerics. Second, we chart out, both
numerically and using adiabatic impulse approximation, the overlap
between the wavefunctions at the end of a drive period $(T=
2\pi/\omega_1$) and the initial ground state wavefunction as a
function of $\omega_1$ and $r$. Our results demonstrate that the
second drive frequency can be used as a tuning parameter to obtain
dynamics induced freezing, {\it i.e.}, a near-perfect overlap
between these wavefunctions. In particular, for the integrable
models studied here, we show that $r=1$ seems to be the best choice
for obtaining dynamic freezing and provide a semi-analytic
explanation of this result. Third, we compute equal-time correlation
functions of the XY model as a function of $\omega_1$ and $r$ and
discuss the signature of dynamics induced freezing in such
correlation functions. Fourth, we study the dynamics of the tilted
Bose-Hubbard model driven out of equilibrium via the two-rate
protocol described above using exact diagonalization (ED) for finite
size systems with $N \le 16$. We show that such a drive leads to
near-perfect dynamics induced freezing for several $\omega_1$ for
$r=2$, provide a qualitative explanation of this phenomenon using a
numerically supported effective two state model, and compute the
dipole-dipole correlation functions which bear the signature of such
freezing phenomenon. Finally, we chart our experiments which can be
used to test our theory.

The plan of the rest of the paper is as follows. In Sec.\
\ref{integra}, we analyze the integrable fermionic Hamiltonian (Eq.\
\ref{fermhamden}) using both adiabatic-impulse approximation and
exact numerics. This is followed by Sec.\ \ref{tiltbh}, where we
study the dynamics of the dipole model (Eq.\ \ref{diham1}). Finally
we summarize our results, discuss their experimental implications,
and conclude in Sec.\ \ref{diss}.

\section {Integrable models}
\label{integra}

In this section we shall study the dynamics of the free fermion
Hamiltonian given by Eq.\ \ref{fermhamden} subjected to the two rate
periodic protocol. We first analyze such dynamics using an
adiabatic-impulse approximation in Sec.\ \ref{ad-imp}. This is
followed by discussion of results from exact numerics in Sec.\
\ref{intres}, carried out using appropriate parameters for 1D XY
model (Eqs.\ \ref{xyham1} and \ref{mapxy}), and their comparison
with those obtained from adiabatic-impulse approximation developed
in Sec.\ \ref{ad-imp}.

\subsection{Adiabatic-Impulse Approximation}
\label{ad-imp}

The basic assumption behind the adiabatic-impulse approximation lies
in dividing the dynamics of a system subjected to a drive into two
distinct regimes \cite{nori1}. The first is the adiabatic regime
where the rate at which the system Hamiltonian changes is small
compared to the instantaneous energy gap; here the dynamics merely
lead to accumulation of a phase of the system wavefunction. The
second constitute the impulse regime where the rate of change of the
Hamiltonian parameter is comparable to or larger than the
instantaneous energy gap; in this regime, excitations are produced
since the system can no longer follow the instantaneous ground
state. In the context of the Hamiltonian given by Eq.\
\ref{fermhamden}, the latter region occurs when the system reaches
the critical point. This approximation therefore becomes accurate
for low-frequency drives where the impulse regime occurs near the
critical point of the Hamiltonian. In what follows, we shall proceed
with the assumption that $H_{\vec k}$ (Eq.\ \ref{fermhamden}) has an
isolated critical point which the system crosses during the
evolution; this assumption fails when the $H_{\vec k}$ has a gapless
region such as for the Kitaev model \cite{kit1,feng1,sen1}. In this
case, the adiabatic-impulse approximation can not be used in its
present form. In this section, we shall only discuss cases where
$H_{\vec k}$, given by Eq.\ \ref{fermhamden}, does not have such
gapless regions.

\begin{figure}
\includegraphics[height=20mm,width=75mm]{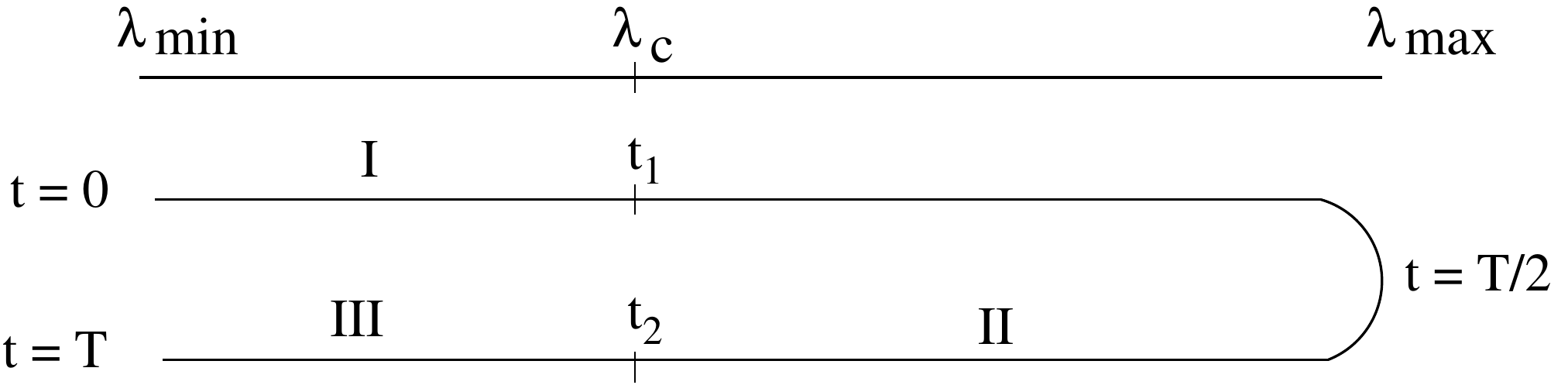}
\caption{Schematic representation of the time-evolution of the
system with a critical point being located at $\lambda= \lambda_c$,
Regions I, II and III are the adiabatic regimes separated by impulse
regions at $t=t_1 \equiv t_{1 \vec k}$ and $t=t_2 \equiv t_{2 \vec
k}$ when the system traverse the critical point. See text for
details.} \label{fig1}
\end{figure}

To treat the dynamics of Eq.\ \ref{fermhamden} subjected to a
two-rate periodic protocol using the adiabatic-impulse
approximation, we first sketch the adiabatic and the impulse regions
in Fig.\ \ref{fig1}. We denote the instantaneous ground state at a
given momentum $\vec k$ to be the ground sate of the $H_{\vec
k}(t)$(Eq.\ \ref{fermhamden}): $|\psi_{\vec k}\rangle_1 = (u_{\vec
k}^0(t), v_{\vec k}^0(t))^T$, where
\begin{eqnarray}
u_{\vec k}^0(t) &=& \frac{ \lambda_2(t)\Delta_{\vec k}}{D_{\vec
k}(t)}, \quad v_{\vec k}^0(t) = -\frac{(\lambda_1(t)-b_{\vec
k})+E_{\vec
k}(t)}{D_{\vec k}(t)} \nonumber\\
E_{\vec k}(t) &=& \sqrt{(\lambda_1(t)-b_{\vec k})^2 +
\lambda_2^2(t)\Delta_{\vec k}^2} \nonumber\\
D_{\vec k}(t) &=& \sqrt{(\lambda_1(t)-b_{\vec k} +E_{\vec k}(t))^2 +
\lambda_2^2(t)\Delta_{\vec k}^2} \label{initwav1}
\end{eqnarray}
The corresponding instantaneous excited state wavefunction for  any
given $\vec k$ is given by $|\psi_{\vec k}(t)\rangle_2= (-v_{\vec
k}^{0}(t), u_{\vec k}^0(t))^T$. Using Eq.\ \ref{initwav1}, one can
write down the wavefunction of the driven system at any time $t$ and
for a given $\vec k$ as
\begin{eqnarray}
|\psi_{\vec k}(t) \rangle &=& b'_{1\vec k}(t) |\psi_{\vec
k}(t)\rangle_1 + b'_{2 \vec k}(t) |\psi_{\vec k}(t)\rangle_2
\label{wavtim1}
\end{eqnarray}
with $(b'_{1 \vec k}(0), b'_{2 \vec k}(0)) = (1,0)$. In what
follows, we shall use the notation $u_{\vec k}(0)=u_{\vec k}^0(T)
\equiv  u_{\vec k}^0$ and $v_{\vec k}(0)=v_{\vec k}^0(T) \equiv
v_{\vec k}^0$.

To obtain expressions for $b'_{1,2 \vec k}(t)$, we first identify
the times $t_{1 \vec k}$ and $t_{2 \vec k}$ (Fig.\ \ref{fig1}) at
which the system comes to an avoided level crossing for a given
momenta $\vec k$. To this end, we note that the instantaneous energy
gap of the system is given by
\begin{eqnarray}
\delta E_{\vec k} &=& 2\sqrt{(\lambda_1(\omega_1 t)-b_{\vec k})^2 +
\lambda_2^2(r \omega_1 t)\Delta_{\vec k}^2}. \label{instengap}
\end{eqnarray}
This gap reaches a minima at $t=t_{1,2 \vec k}$ at an avoided level
crossing for which $\partial \delta E_{\vec k}/\partial t =0$. This
leads to
\begin{eqnarray}
\dot \lambda_1(\omega_1 t_{1\vec k}) \left[ \lambda_1(\omega_1 t_{1
\vec k}) - b_{\vec k} \right] + \dot \lambda_2(\omega_2 t_{1\vec k})
\lambda_2(\omega_2 t_{1\vec k}) \Delta_{\vec k}^2 &=& 0,
\label{tikeq} \nonumber\\
\end{eqnarray}
where $\dot x = \partial_t x$. For simple enough protocols, it is
possible to solve this equation to obtain an analytic expression for
$t_{1,2 \vec k}$. For example if $\lambda_1 = \lambda_2= A
\cos(\omega_1 t)$, one obtains $ \omega_1 t_{1\vec k} = \arccos[A
b_{\vec k}/(A^2 + \Delta_{\vec k}^2)]$ and $ \omega_1 t_{2 \vec k} =
2\pi - \omega_1 t_{1 \vec k}$. In other, more complicated cases, one
needs to solve Eq.\ \ref{tikeq} numerically to obtain $t_{1,2 \vec
k}$.

Let us now consider the evolution of the system in region I (Fig. \
\ref{fig1}) for $t < t_{1\vec k}$. In this region, the system is in
the adiabatic regime and the wavefunction merely gathers a kinetic
phase during evolution. The evolution operator in this regime is
thus given by \cite{nori1}
\begin{eqnarray}
U_1(t,0) &=& e^{ - i \Omega_{1 \vec k}(t) \tau_3}, \,\, \Omega_{1
\vec k}(t) = \frac{1}{2} \int_0^t \delta E_{\vec k}(t') dt'
\label{evo1}
\end{eqnarray}
Similarly, in region II for $ t_{1 \vec k} < t < t_{2 \vec k}$ and
III, for $ t_{2 \vec k} < t \le T$, the evolution operators are
given by
\begin{eqnarray}
U_2(t,t_{1 \vec k}) &=& e^{ - i \Omega_{2 \vec k}(t) \tau_3}, \,\,
\Omega_{2 \vec k}(t) = \frac{1}{2} \int_{t_{1\vec k}}^t  \delta
E_{\vec k}(t') dt' \label{evo2} \nonumber\\
U_3(t,t_{2 \vec k}) &=&
e^{ - i \Omega_{3 \vec k}(t) \tau_3}, \,\, \Omega_{3 \vec k}(t) =
\frac{1}{2} \int_{t_{2\vec k}}^t  \delta E_{\vec k}(t') dt'
\label{evo3} \nonumber\\
\end{eqnarray}

Around $t=t_{1,2 \vec k}$, the system enters the impulse regions and
excitations are produced. The key aspect of the adiabatic-impulse
approximation is to approximate the impulse region to be exactly at
$t_{1,2, \vec k}$. Near these points, the effective Hamiltonian has
the form
\begin{eqnarray}
{\mathcal H}^{a}_{\vec k} (t) &=& \left[ \Lambda_{1 \vec k} + \dot
\lambda_1(\omega_1 t_{a \vec k}) \delta t_{a \vec k} \right] \tau_3
\nonumber\\
&& +\left[\Lambda_{2 \vec k} + \dot \lambda_2 (\omega_2 t_{a \vec
k})
\delta t_{a \vec k} \Delta_{\vec k} \right] \tau_1  \nonumber\\
\Lambda_{1 \vec k} &=& \lambda_1(\omega_1 t_{a \vec k}) - b_{\vec k}
\nonumber\\
\Lambda_{2 \vec k} &=& \Delta_{\vec k} \lambda_2(\omega_2 t_{a \vec
k}), \quad \delta t_{a \vec k} = t- t_{a \vec k}, \label{impham1}
\end{eqnarray}
where $a=1,2$ for the two impulse regions. We note that in contrast
to standard applications \cite{nori1, ks3}, the impulse Hamiltonian
${\mathcal H}_{\vec k}$  for the two-rate protocol is not of the
Landau-Zener form; in contrast, it has time-dependence both in the
diagonal and off-diagonal terms. To apply the Landau-Zener result,
we therefore cast it in the standard form following the method
suggested in Ref.\ \onlinecite{jay1}. To this end, we first write
${\mathcal H}_{\vec k}(t)$ in terms of a new set of Pauli matrices
as
\begin{eqnarray}
{\mathcal H}^{a}_{\vec k} (t) &=& \mu_{1 \vec k}^a (\delta t_{a \vec
k} - t'_{a \vec k}) \tilde \tau_3 + \mu_{2 \vec k}^a \tilde \tau_1
\label{impham2}
\end{eqnarray}
A comparison of Eqs.\ \ref{impham1} and \ref{impham2} shows that for
both equations to be simultaneously correct, one needs
\begin{eqnarray}
\tilde \tau_3 \mu_{1 \vec k}^a &=&  \dot \lambda_1(\omega_1 t_{a
\vec
k}) \tau_3 + \dot \lambda_2(\omega_2 t_{a \vec k}) \Delta_{\vec k} \tau_1  \nonumber\\
\tilde \tau_1 \mu_{2 \vec k}^a &=&  (\Lambda_{1 \vec k} + \dot
\lambda_1(\omega_1 t_{a \vec k}) t'_{a \vec k}) \tau_3 \nonumber\\
&& + (\Lambda_{2 \vec k} + \dot \lambda_2(\omega_2 t_{a \vec k})
t'_{a \vec k} \Delta_{\vec k}) \tau_1 \label{eqrel1}
\end{eqnarray}
The identities $ ({\rm Det} [\tilde \tau_{1,3}])^2 =1$ and $\{\tilde
\tau_3,\tilde \tau_1\}=0$, then leads to
\begin{eqnarray}
\mu_{1 \vec k}^a&=& \sqrt{\dot \lambda^2_1(\omega_1 t_{a \vec k}) +
\dot
\lambda^2_2(\omega_2 t_{a \vec k}) \Delta^2_{\vec k}} \nonumber\\
\mu_{2 \vec k}^a &=& \left[ (\Lambda_{1 \vec k} + \dot
\lambda_1(\omega_1 t_{a \vec k}) t'_{a \vec k})^2 \right.
\nonumber\\
&& \left. + (\Lambda_{2 \vec k} + \dot \lambda_2(\omega_2 t_{a \vec
k}) t'_{a \vec k}
\Delta_{\vec k})^2 \right]^{1/2} \nonumber \\
t'_{a \vec k} &=& - (\dot \lambda_1(\omega_1 t_{a \vec k})\Lambda_{1
\vec k}  + \dot \lambda_2(\omega_2 t_{a \vec k})\Lambda_{2 \vec
k})/\mu^2_{1 \vec k}. \label{eqrel2}
\end{eqnarray}

For the Hamiltonian given by Eqs.\ \ref{impham2} and \ref{eqrel2},
the probability of excitation production can be read off as
\cite{pol1}
\begin{eqnarray}
p_{\vec k}^a &=& \exp[-2 \pi \delta_{\vec k}^a], \quad \delta_{\vec
k}^a = |\mu_{2 \vec k}^a|^2/\left(2 |\mu_{1 \vec k}^a|\right)
\label{kz1}
\end{eqnarray}
Using Eq.\ \ref{kz1}, one can write down the unitary evolution
matrices $\Sigma_a$ which propagates the wavefunction across $t=t_{a
\vec k}$ \cite{nori1}. The elements of $\Sigma_a$ can be expressed
in terms of $p_{\vec k}^a$ and the Stokes phase $\tilde \phi_{s \vec
k}^a$ as
\begin{eqnarray}
\Sigma_a &=& \left( \begin {array} {cc}  \sqrt{1-p^a_{\vec k}}e^{-i
\tilde \phi_{s\vec k}^a} & - \sqrt{p_{\vec k}^a} \\ \sqrt{p_{\vec
k}^a} & \sqrt{1-p^a_{\vec k}}e^{i \tilde
\phi_{s\vec k}^a} \end {array} \right), \label{evo4} \\
\tilde \phi_{s \vec k}^a &=& -\frac{\pi}{4} + \delta_{\vec k}^a
\left( \ln \delta_{\vec k}^a -1 \right) + {\rm Arg} \Gamma \left(1 -
i \delta_{\vec k}^a \right) \nonumber
\end{eqnarray}
where $\Gamma$ denotes the gamma function. Using Eqs.\ \ref{evo1},
\ref{evo3}, and \ref{evo4}, we finally obtain
\begin{eqnarray}
\left(\begin{array} {c} b'_{1\vec k} \\ b'_{2 \vec k}
\end{array}\right) &=& U_{3}( T,t_{2 \vec k}) \, \Sigma_{\vec k}^2 \, U_{2}(
t_{2 \vec k}, t_{1\vec k}) \, \Sigma_{\vec k}^1 \, U_{1}( t_{1 \vec
k},0) \left(\begin{array} {c} 1 \\ 0 \end{array} \right)
\nonumber\\
&=& \left( \begin{array}{cc} q_{+\vec k} & -q_{- \vec k}^{\ast} \\
q_{- \vec k} & q_{+\vec k}^{\ast} \end{array} \right)
\left(\begin{array} {c} 1 \\ 0 \end{array} \right) =
\left(\begin{array} {c} q_{+ \vec k} \\ q_{- \vec k} \end{array}
\right) \nonumber\\
q_{+\vec k} &=& e^{-i \Omega_{3 \vec k}} \left[
\sqrt{(1-p_{\vec k}^1)(1-p_{\vec k}^2)} e^{-i \sum_{a=1}^2
(\Omega_{a \vec k} + \tilde \phi_{s \vec k}^a)} \right.
\nonumber\\
&& \left. - \sqrt{p_{\vec k}^1 p_{\vec k}^2} e^{i(\Omega_{2 \vec k}-
\Omega_{1 \vec k})} \right] \nonumber\
\end{eqnarray}
\begin{eqnarray}
q_{- \vec k} &=& e^{i \Omega_{3 \vec k}} \left[ \sqrt{(1-p_{\vec
k}^2)p_{\vec k}^1} e^{ i (\tilde \phi_{s \vec k}^2 + \Omega_{2 \vec
k} - \Omega_{1 \vec k})} \right. \nonumber\\
&& \left. + \sqrt{(1-p_{\vec k}^1)p_{\vec k}^2} e^{ -i (\tilde
\phi_{s \vec k}^1 + \Omega_{1 \vec k} + \Omega_{2 \vec k})} \right]
\label{adimpfinaleq} \\
|\psi_{\vec k}(T)\rangle &=& \left(\begin{array} {c} q_{+\vec k} u_{\vec k}^0-q_{-\vec k} v_{\vec k}^0 \\
q_{+\vec k} v_{\vec k}^0 + q_{-\vec k} u_{\vec k}^0\end{array}\right) = \left(\begin{array} {c} u_{\vec k}(T)\\
v_{\vec k}(T)
\end{array}\right) \nonumber\
\end{eqnarray}

Eq.\ \ref{adimpfinaleq} is the central result of this section. It
allows us to express the final wavefunction at the end of a drive
period for an arbitrary two-rate drive protocol. We note that the
expressions of $q_{\pm \vec k}$ obtained in Eq.\ \ref{adimpfinaleq}
constitutes a generalization of its single-rate protocol counterpart
\cite{nori1}. Furthermore, it allows us to obtain semi-analytic
expressions for excitation density and all fermionic correlators at
the end of the drive. In the context of XY model in a transverse
field where the transverse spin correlators can be expressed in
terms of fermionic correlators \cite{subir1}, Eq.\
\ref{adimpfinaleq} allows us to obtain analytic expressions for the
transverse magnetization and other relevant spin correlation
functions. In what follows, we shall be interested in equal-time
fermionic correlation functions
\begin{eqnarray}
C_{{\vec i}- {\vec j}} &=&  L^{-d} \sum_{\vec k} \langle \psi_{\vec
k}(T)| (c_{\vec k}^{\dagger} c_{\vec k} + {\rm h.c.}) |\psi_{\vec
k}(T) \rangle \nonumber\\
&& \times \cos[\vec k \cdot (\vec i - \vec j)] \nonumber\\
&=&  L^{-d} \sum_{\vec k} |v_{\vec k}(T)|^2 \cos[\vec k \cdot (\vec i - \vec j)] \label{correq1} \\
F_{{\vec i }-{\vec j}} &=&  L^{-d} \sum_{\vec k} \langle \psi_{\vec
k}(T)|(c_{k} c_{-k} + {\rm h.c.}) |\psi_{\vec k}(T)\rangle
\nonumber\\
&& \times \sin[\vec
k \cdot(\vec i - \vec j)] \nonumber\\
&=&  L^{-d} \sum_{\vec k} (u^{\ast}_{\vec k}(T) v_{\vec k}(T) + {\rm
h.c.}) \sin[\vec k \cdot (\vec i - \vec j)] \nonumber
\end{eqnarray}
and the excitation density $n_d$
\begin{eqnarray}
n_d = 1-F, \quad F = L^{-d} \sum_{\vec k} |\langle \psi_{\vec
k}^0|\psi_{\vec k}(T) \rangle|^2 \label{wavov}
\end{eqnarray}
where $F$ is the wavefunction overlap at the end of the drive which
approaches close to unity for dynamics induced freezing, the sum
over momenta extends over half of the Brillouin zone, and $L$ is the
linear dimension of the system. These quantities can be expressed in
terms of $q_{\pm \vec k}$ and the instantaneous eignstates at $t=T$,
$u_{\vec k}^0$ and $v_{\vec k}^0$, using Eqs. \ref{adimpfinaleq} and
\ref{wavtim1}, as
\begin{eqnarray}
n_d &=& 1-F = L^{-d} \sum_{\vec k} |q_{- \vec k}|^2
\label{wavovadimp} \\
C_{{\vec i}- {\vec j}} &=&  L^{-d} \sum_{\vec k} |q_{- \vec k}
u_{\vec k}^0 + q_{+ \vec k} v_{\vec k}^0|^2
\cos[\vec k \cdot (\vec i - \vec j)] \nonumber\\
F_{{\vec i }-{\vec j}} &=&  L^{-d} \sum_{\vec k} \Big[ \left( q_{+
\vec k}^{\ast} u_{\vec k}^0 - q_{-\vec k}^{\ast} v_{\vec k}^0
\right) \left(q_{- \vec k} u_{\vec k}^0 + q_{+ \vec k} v_{\vec k}^0
\right) \nonumber\\
&& + {\rm h.c.} \Big] \sin[\vec k \cdot (\vec i - \vec j)]
\label{correqadimp}
\end{eqnarray}

In the next section, we shall use Eq.\ \ref{correqadimp} to discuss
the behavior of magnetization, excitation density and correlation
functions of the transverse field XY models as computed using
adiabatic-impulse technique and compare with the exact numerics.

\begin{figure}
\includegraphics[height=60mm,width=\linewidth]{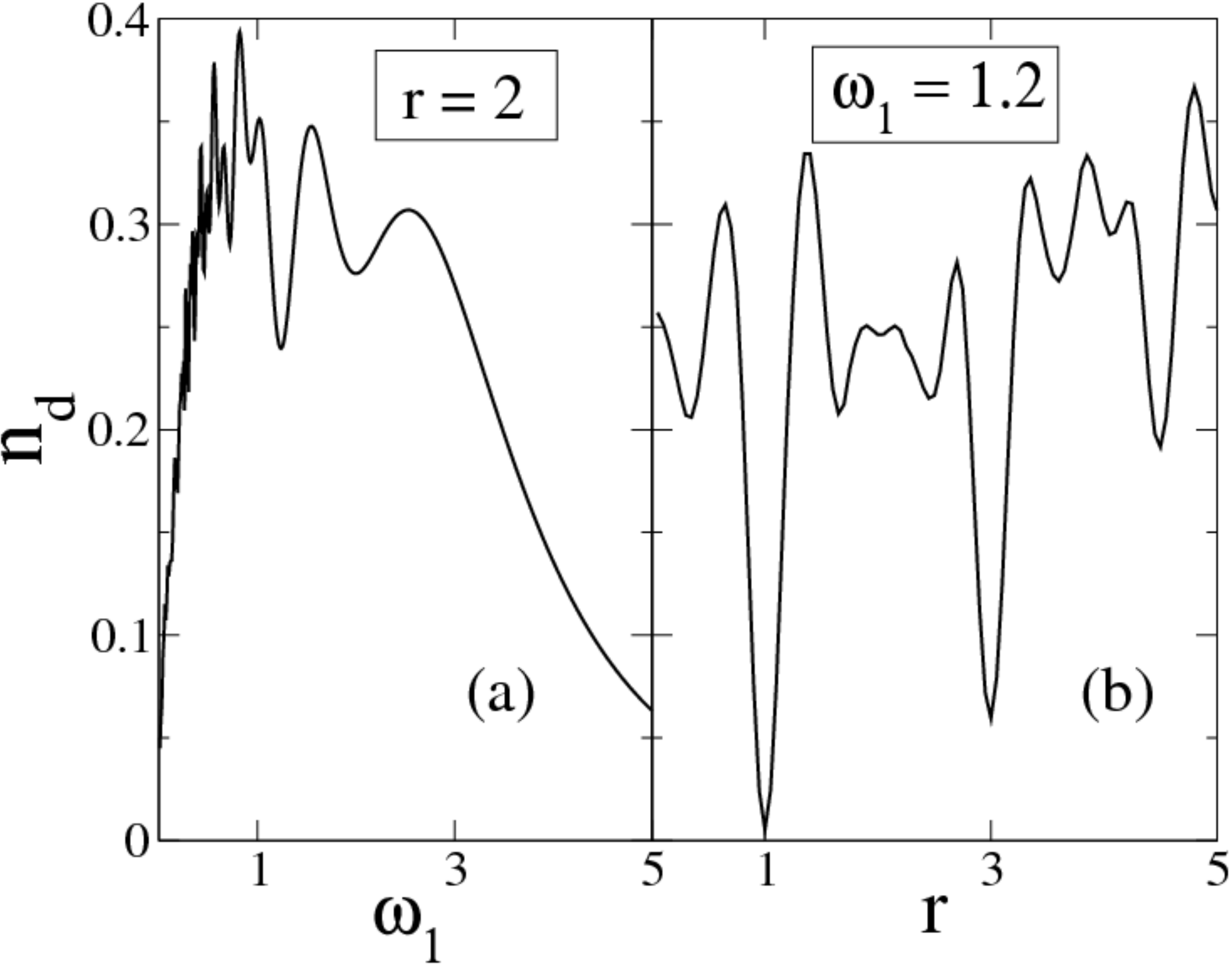}
\caption{(a) Plot of the defect density $n_d$ of the 1D XY model in
a transverse field as a function of the drive frequency $\omega_1$
for $r=2$. Here we have chosen $J_x+J_y=1$, $J_x-J_y = \cos(r
\omega_1 t+ \phi)$, and $h= A\cos(\omega_1 t)$ with $A=4$ and
$\phi=0$. (b) Plot of $n_d$ as a function of $r$ for $\omega_1=1.2$
and with all parameters same is (a). Note that $r=0$ correspond to
the single parameter drive protocol. A clear signature of dynamic
freezing is found at $r=1$ and $\omega_1=1.2$.} \label{fig2}
\end{figure}

\subsection{Results for integrable models}
\label{intres}

In this section, we discuss numerical results obtained by exact
numerical solution of the driven Hamiltonian Eq.\ \ref{fermhamden}
and compare those obtained Sec.\ \ref{ad-imp} using generalized
adiabatic-impulse approximation. To do this, we take advantage of
the fact that $H_k$ represents a two-level Hamiltonian for each
$(\vec k, -\vec k)$ pair. Thus for every given $\vec k$, it is
possible to write a wavefunction $|\psi_{\vec k}(t)\rangle =
(u_{\vec k}(t), v_{\vec k}(t))^T$ which satisfies
\begin{eqnarray}
i \partial_t u_{\vec k}(t) &=& (\lambda_1(t)-b_{\vec k}) u_{\vec
k}(t) + \lambda_2(t) \Delta_{\vec k} v_{\vec
k}(t) \nonumber\\
i \partial_t v_{\vec k}(t) &=& -(\lambda_1(t)-b_{\vec k}) v_{\vec
k}(t) + \lambda_2(t) \Delta_{\vec k} u_{\vec k}(t), \label{numeq1}
\end{eqnarray}
where we first choose the parameters $\lambda_{1(2)}(t)$, $b_{\vec
k}$, and $\Delta_{\vec k}$ to be those for XY model in a transverse
field (Eq.\ \ref{mapxy}) with $\lambda_1 = A \cos(\omega_1 t)$,
$\lambda_2= \cos(r \omega_1 t +\phi)$, and $J_{x(y)}= (1
+(-)\lambda_2(t))/2$. Note that our choice of parameters is
equivalent to scaling all energy parameters of the Hamiltonian with
$J_x+J_y$ and all time scales with $\hbar/(J_{x}+J_y)$. A numerical
solution of Eq.\ \ref{numeq1} enables us to obtain $
|\psi_k(t)\rangle$ which leads to the defect density $n_d$ and the
correlation functions $C_{ij}$ and $F_{ij}$ using Eqs.\ \ref{wavov}
and \ref{correq1} at the end of a single drive period $T=
2\pi/\omega_1$.

\begin{figure}
\includegraphics[height=60mm,width=\linewidth]{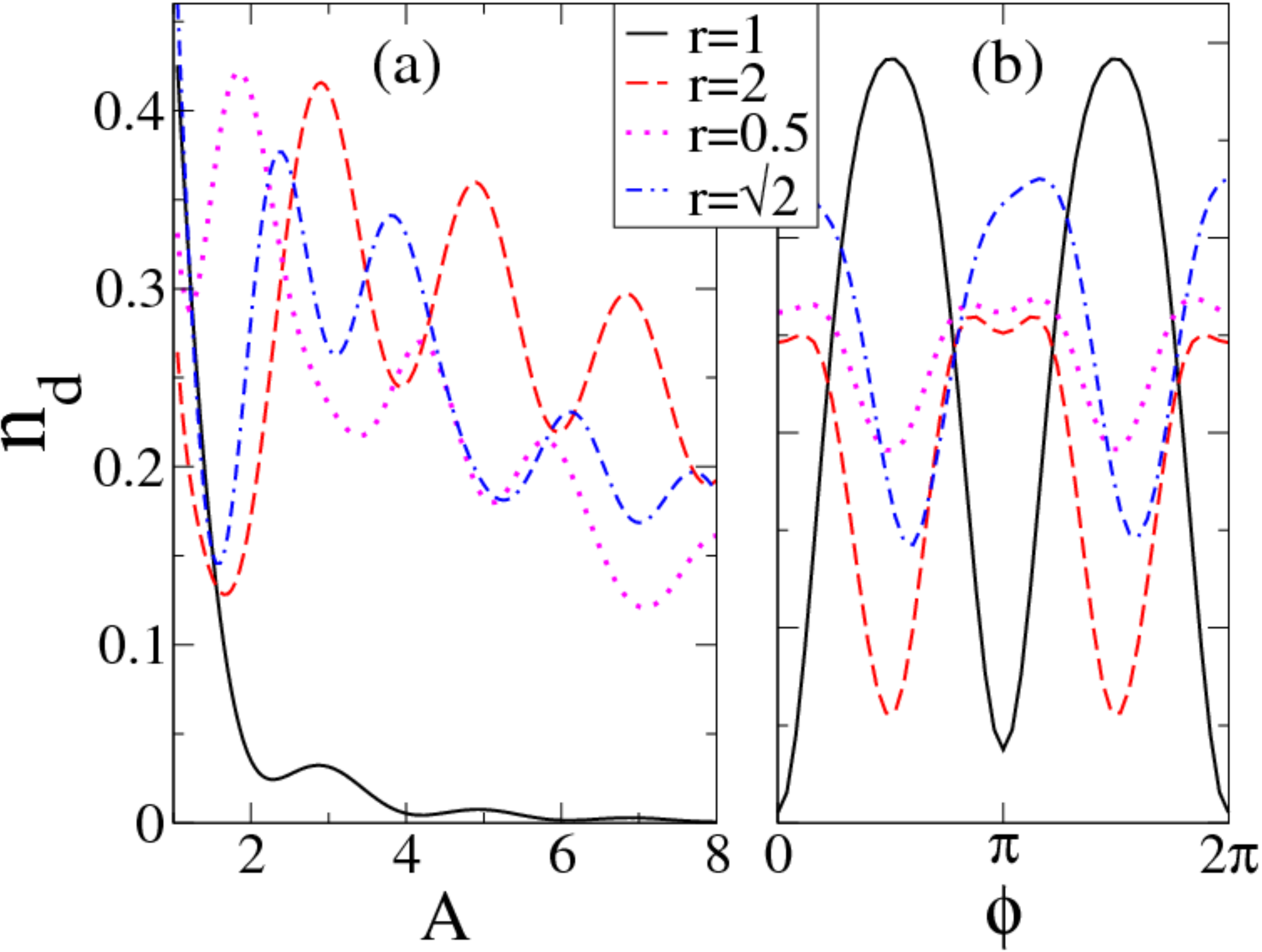}
\caption{(a) Variation of the defect density $n_d$ as a function of
the drive amplitude for $\omega_1=1.2$, $\phi=0$, and several
representative values of $r$. All other parameters are same in Fig.\
\ref{fig2}. (b) Variation of $n_d$ as a function of the initial
phase $\phi$ of the drive for $A=4$. } \label{fig3}
\end{figure}

\begin{figure}
\includegraphics[height=60mm,width=\linewidth]{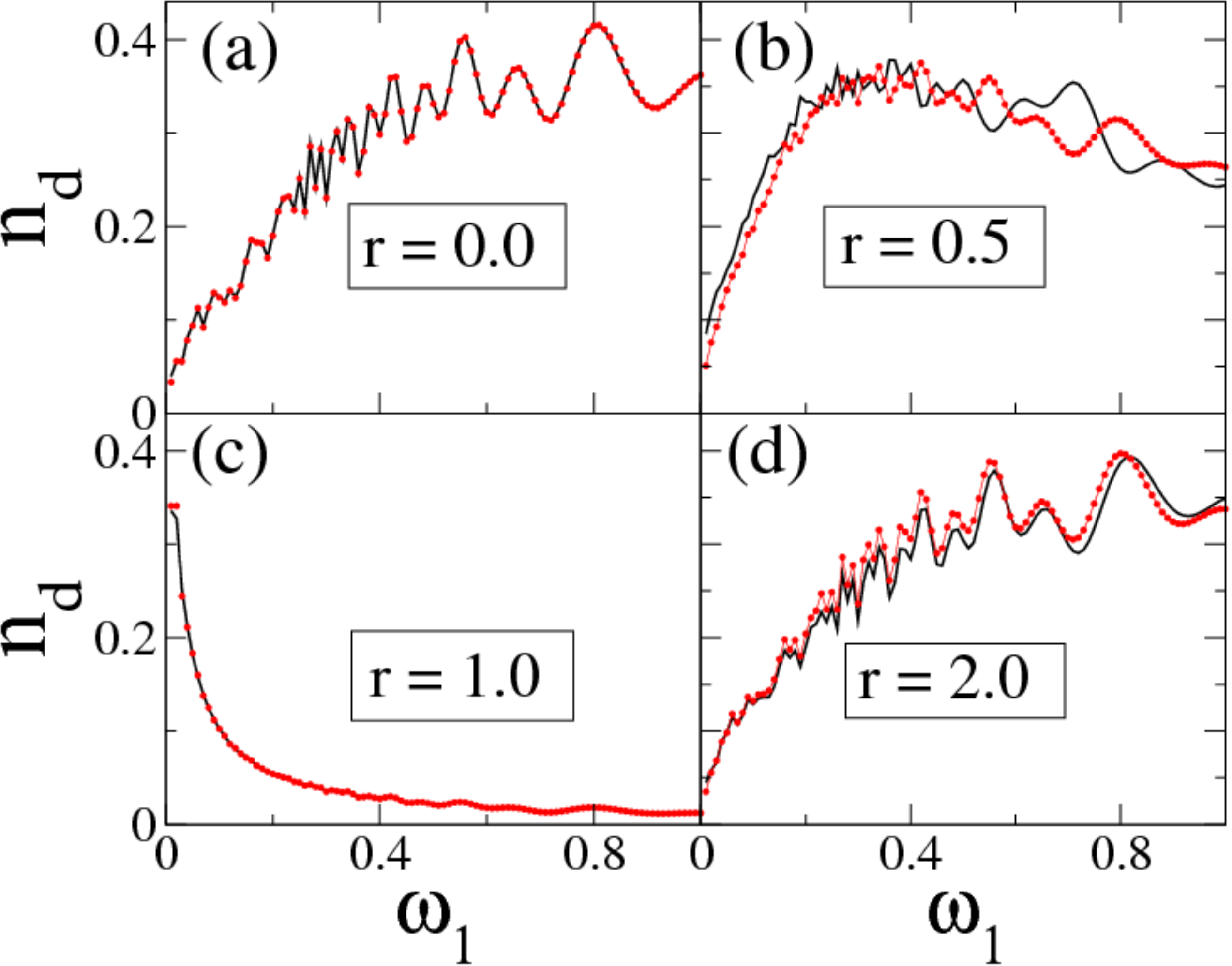}
\caption{Comparison of exact numerical result (black solid line) for
variation of $n_d$ as a function of $\omega_1$ with that obtained
from the adiabatic-impulse method (red dotted line) detailed in
Sec.\ \ref{ad-imp} for (a) $r=0$, (b) r=0.5, (c) $r=1$ and (d)
$r=2$. All parameters are same as in Fig.\ \ref{fig2}.} \label{fig4}
\end{figure}

The results of the numerical procedure described above is shown in
Figs.\ \ref{fig2}, \ref{fig3}, and \ref{fig4}. In Fig.\
\ref{fig2}(a), we plot $n_d$ as a function of $\omega_1$ for $r=2$
while in Fig.\ \ref{fig2}(b), we plot $n_d$ as a function of $r$ for
$\omega_1= 1.2$. These plots clearly demonstrate that $n_d$
approaches zero close to $r=1$ and $ \omega_1=1.2$; at this point,
$|\psi_{\vec k} (T)\rangle$ has almost perfect overlap with
$|\psi_{k}(0)\rangle$ for all $k$ which signifies dynamic freezing.
We note that such a freezing does not occur for a single parameter
drive protocol (which corresponds to $r=0$) for any $\omega_1$; thus
our results demonstrate that the second drive frequency, or
equivalently $r$, for a two-parameter drive protocol can act as a
tuning parameter to obtain such freezing. A plot of $n_d$ as a
function of the drive amplitude $A$ (Fig\ \ref{fig3}(a)) and the
initial phase $\phi$ (Fig\ \ref{fig3}(b)) for several representative
values of $r$ indicates that $n_d$ indeed reaches a minimum of $r=1$
for all $A>1$; $n_d$ decreases rapidly with increasing $A$ for $r=1$
while it displays a slow oscillatory decay for other values of $r$.
Further $n_d$ reaches a minimal value for $\phi=0$ as shown in Fig.\
\ref{fig3}(b). A comparison of these exact numerical results with
those obtained using adiabatic-impulse approximation (Eq.\
\ref{wavovadimp}) for $n_d$ is shown in Fig.\ \ref{fig4}. We find
that the adiabatic-impulse approximation turns out to match the
exact numerical results for $n_d$ for all $\omega_1 \le 1$ and for
any $r$.

From Figs.\ \ref{fig2}, \ref{fig3}, and \ref{fig4}, it becomes
apparent that the dynamics of $r=1$ is qualitatively different from
that at $r \ne 1$ since it shows a rapid decay of $n$ with both
$\omega_1$ and $A$. We now present a reason for this behavior. We
first note that at $r=1$, the dynamics is controlled by $H_{\vec
k}^{r=1}$ given by
\begin{eqnarray}
H_{\vec k}^{r=1} = \left(A f(\omega_1 t) -b_{\vec k}\right) \tau_3 +
\Delta_{\vec k} f(\omega_1 t) \tau_1, \label{r1eq1}
\end{eqnarray}
where we have chosen $f(\omega_1 t) =\cos(\omega_1 t)$ for Figs.\
\ref{fig2}..\ref{fig4}. It turns out that $H_{\vec k}^{r=1}$ can be
cast onto a form \cite{jay1}
\begin{eqnarray}
H_{\vec k}^{r=1} = \alpha_{\vec k}( f(\omega_1 t)-t_{1 \vec k} )
{\tilde \tau_3} + \alpha_{2 \vec k} {\tilde \tau_1} \label{r1eq2}
\end{eqnarray}
where ${\tilde \tau_{1,3}}$ are new Pauli matrices which are related
to $\tau_{1,3}$ by
\begin{eqnarray}
\alpha_{1\vec k} {\tilde \tau_3} &=& A \tau_3 + \Delta_{\vec k}
\tau_1 \nonumber\\
\alpha_{2 \vec k} {\tilde \tau_1} &=& (A t_{1\vec k} - b_{\vec k})
\tau_3 + \Delta_{\vec k} t_{1\vec k}  \tau_1. \label{r1eq3}
\end{eqnarray}
Using the properties that ${\rm Det}{\tilde \tau_{1,3}}=-1$ and
$\{\tau_3, \tau_1\}_{+}=0$ (where $\{..\}_+$ denotes anticommutator)
one obtains
\begin{eqnarray}
\alpha_{1 \vec k} &=& -\sqrt{A^2 + \Delta_{\vec k}^2} \nonumber\\
t_{1 \vec k} &=& A b_{\vec k}/\alpha_{1 \vec k }^2, \quad \alpha_{2
\vec k} = -\Delta_{\vec k} b_{\vec k}/\alpha_{1 \vec k}
\label{r1eq4}
\end{eqnarray}

We note that such a transformation allows us to write $H_{\vec k}$
in terms of new Pauli matrices and with all time dependence being
shifted to the diagonal term; such transformation exist for
non-linear drive protocols only for $r=1$, {\it i.e.}, when the
off-diagonal and the diagonal terms of $H_{\vec k}^{r=1}$ are driven
by identical drive protocol. The avoided level crossings occur at
$t=t_{0\vec k}$ and $t= T-t_{0\vec k}$ where $f(\omega_1 t_0)
=b_{\vec k}$. Linearizing $H_{\vec k}^{r=1}$ around these point, we
find that within adiabatic-impulse approximation, the  Landau-Zener
transition probability $p_{\vec k}$ can be read-off from Eq.\
\ref{r1eq2} to be
\begin{eqnarray}
p_{\vec k} &=& e^{-\pi|\alpha_{2 \vec k}|^2/|\alpha_{1\vec k} \dot
f(\omega_1 t_{0 \vec k})|} \label{pkeq1}
\end{eqnarray}
We note that the maximal excitation production occurs from modes
around $\vec k=\vec k_0$. Assuming that $\Delta_{\vec k} \sim |\vec
k -\vec k_0|^z$ around $\vec k_0$ and $b_{\vec k_0} \ne 0$, we find
that for $f_1(\omega_1 t) = \cos(\omega_1 t)$
\begin{eqnarray}
p_{\vec k} \simeq e^{-b_{\vec k}^2 |\vec k -\vec k_0|^{2
z}/(\omega_1 A^3 |\sqrt{1-b_{\vec k}^2/A^2}|)} \label{pkeq2}
\end{eqnarray}
From Eq.\ \ref{pkeq2} we find that $p_k \to 1$ as $A$ or $\omega_1$
increases; thus the excitation probability after one cycle of the
drive $P_{\vec k} \sim p_{\vec k} (1-p_{\vec k})$ rapidly decays to
zero with increasing $A$ or $\omega_1$. It is instructive to compare
this decay to the $r=0$ case where the off-diagonal term in
time-independent. In this case one can read off $p_{\vec k}$ using
the standard results provided in Ref.\ \onlinecite{nori1}: $p_{\vec
k}(r=0)= \exp[-|\vec k -\vec k_0|^{2 z}/(\omega_1 A |\sqrt{1-b_{\vec
k}^2/A^2}|)]$. Thus we find $p_{\vec k}$ approaches unity with
increasing drive amplitude much more rapidly for $r=1$ which
explains the almost monotonic decay of $P_{\vec k}$ and hence $n_d$
for $r=1$.

\begin{figure}
\includegraphics[height=70mm,width=\linewidth]{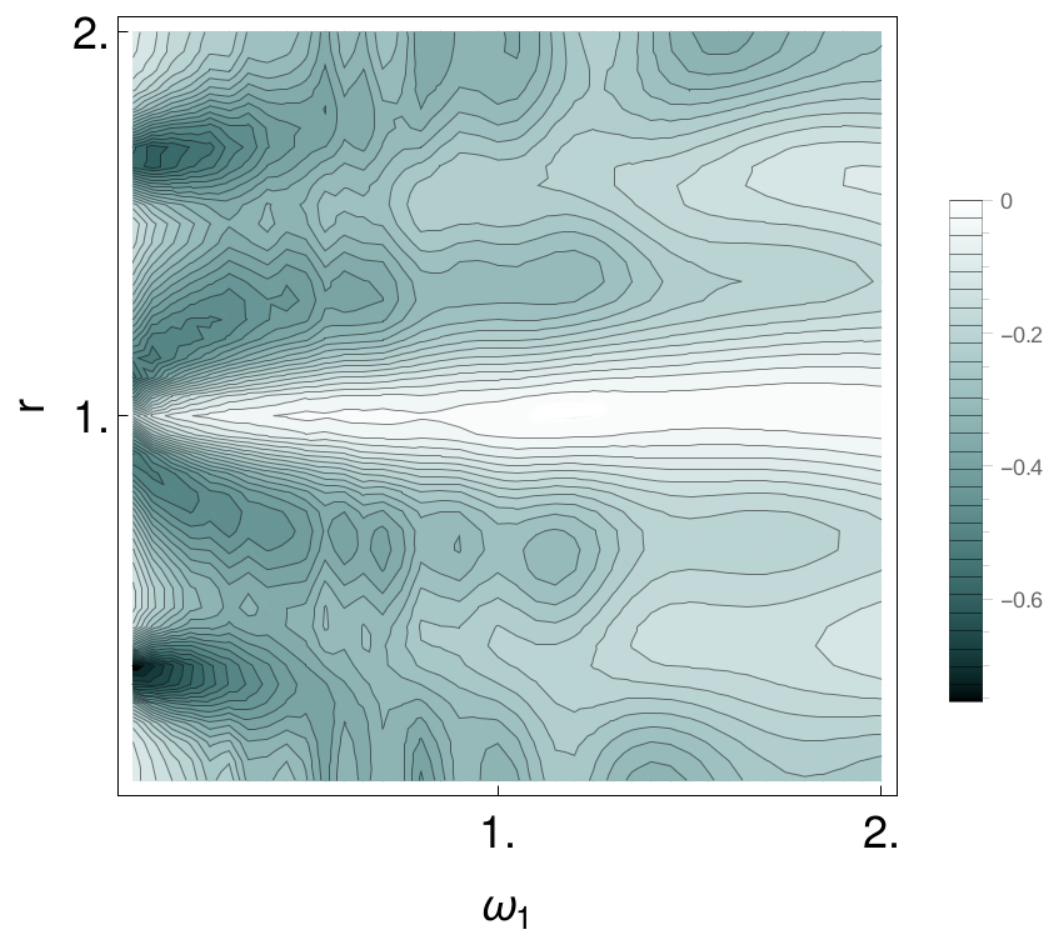}
\caption{Plot of the transverse magnetization $\delta m_z =
m_z(T)-m_z(0)$ as a function of $r$ and $\omega_1$. The dark patch
at $r=1$ is a signature of the dynamical freezing phenomenon which
leads to $\delta m_z =0$. } \label{fig5}
\end{figure}

Finally we plot the change in transverse magnetization at $t=T$,
$\delta m_z(T)= m_z(T)-m_z(0)$ for the XY model as a function of $r$
and $\omega_1$ in Fig.\ \ref{fig5}. We note that this magnetization
can be easily obtained from the expressions of the correlation
function $C_{\vec i -\vec j}$ given in Eq.\ \ref{correq1}
\begin{eqnarray}
m_z (T) &=& 2 L^{-d} \sum_{\vec k} |v_{\vec k}(T)|^2-1 =2C_0(T)-1.
\end{eqnarray}
We note $\delta m_z$ vanishes if dynamic freezing occurs; such a
vanishing of $\delta m_z$ is clearly seen along the $r=1$ line in
Fig.\ \ref{fig5}.

\section {Titled Bose-Hubbard model}
\label{tiltbh}

In this section, we shall study the two-rate periodic drive protocol
for the tilted Bose-Hubbard model whose Hamiltonian is given by Eq.\
\ref{tiltham1}. We shall analyze the dynamics of this model for
$\hbar \omega_1, \hbar \omega_2 \ll U, E$; in this regime, the
dipole Hamiltonian given by Eq.\ \ref{diham1} may be used to
describe the dynamics of the bosons. In what follows, we shall make
the electric field $E$ and the hopping $J$ time dependent according
to the protocol
\begin{eqnarray}
E(t) &=& U - E_0 \cos(\omega_1 t) \nonumber\\
J(t) &=& J_0 \left( 1 + \cos(r \omega_1 t) \right) \label{prot1}
\end{eqnarray}
with $J(0)=1$. The electric field $E$ may be made time-dependent by
varying the Zeeman field used to generate it as a function of time
\cite{bakr1} while the detailed method creating a time dependent $J$
in realistic experimental situation has been discussed in Ref.\
\onlinecite{rs1}. Following the convention used in previous studies
on this model, we shall choose $U/J(0)=40$, $E_0/J(0)=10$ and
$n_0=1$ so that the dynamics takes the system through the critical
point located $E_c= U + 1.83 J$ \cite{subir1}.

\begin{figure}
\includegraphics[height=70mm,width=\linewidth]{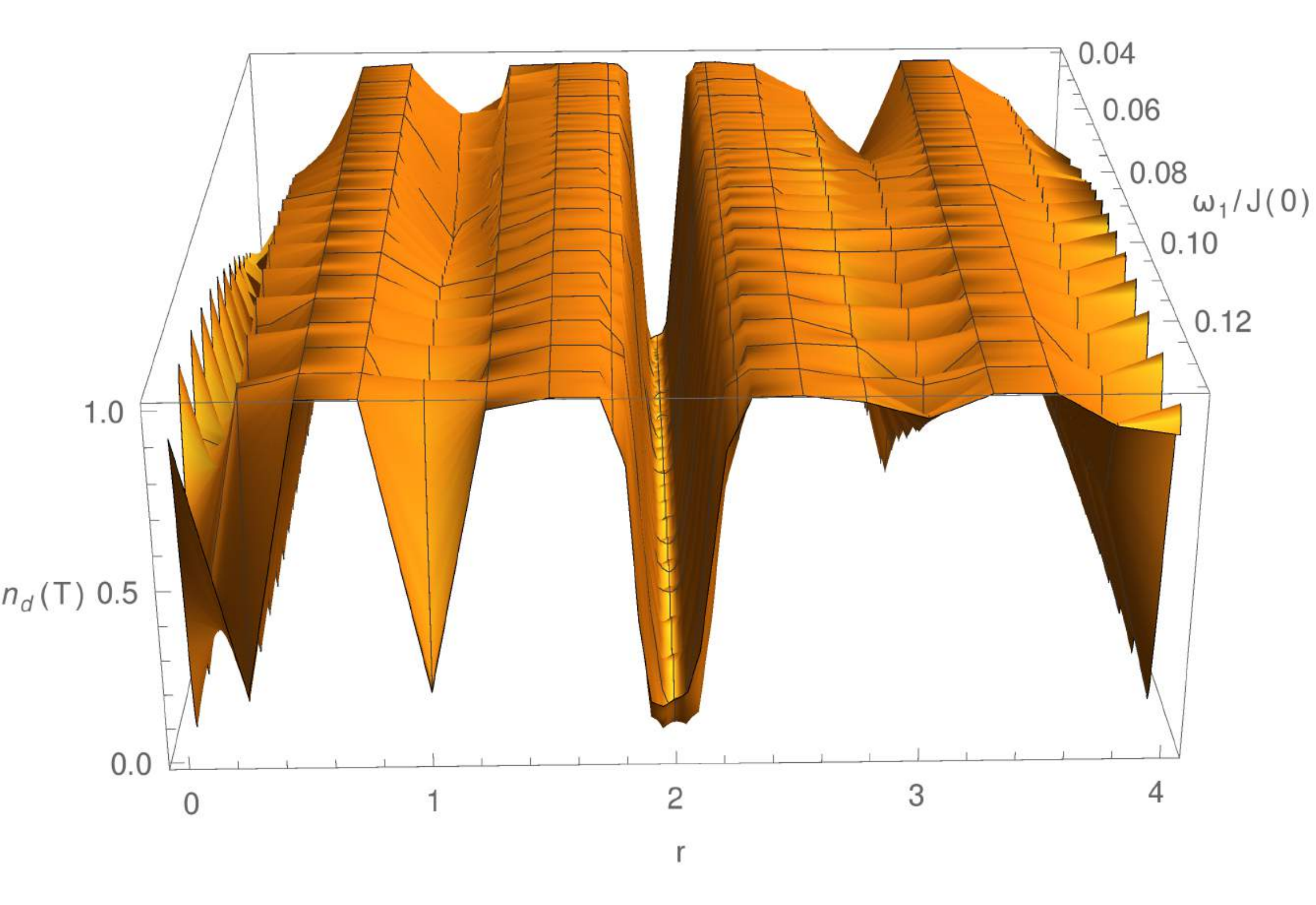}
\caption{Plot of the excitation density $n_d=1-F$ at $t=T$ as a
function of $\omega_1$ and $r$ for $N=12$, $E(0)=30 J(0)$ and $U=40
J(0)$. Note that $n_d \simeq 0$ for $r=2$ independent of
$\omega_1$.} \label{fig6}
\end{figure}

To solve for the dynamics, we obtain the instantaneous eigenvalues
and eigenvectors of the Hamiltonian $H_d$ given by Eq.\ \ref{diham1}
with $J(t)=J(0)$ and $E(t)=E(0)$ using exact diagonalization (ED)
for a bosonic chain with $N\le 16$ sites. We denote these many-body
states by $|m \rangle$; they satisfy $H_d[J(0),E(0)] |m\rangle =
\epsilon_m |m\rangle$ with $\epsilon_1$ and $|1\rangle$ being the
ground state energy and wavefunction respectively. Having obtained
$|m\rangle$, the wavefunction $|\psi(t)\rangle$ at any time which
satisfies the Schrodinger equation $i \hbar \partial_t
|\psi(t)\rangle= H_d[J(t),E(t)] |\psi(t)\rangle$, can be expressed
as
\begin{eqnarray}
|\psi(t) \rangle = \sum_m  c_m (t) |m\rangle. \label{wavexp1}
\end{eqnarray}
To obtain $c_m(t)$, we write the time-dependent Hamiltonian $H_d$ at
any instant as
\begin{eqnarray}
&& H_d[J(t), E(t)] = H_d[J(0),E(0)] - J_0 \cos(r \omega_1 t) \nonumber\\
&& \times \sum_{\ell} (f_{\ell} + f_{\ell}^{\dagger}) + E_0
(1-\cos(\omega_1 t))
\sum_{\ell} n_{\ell} \label{wavexp2} \\
&=& H_d[J(0),E(0)] -J_0 \cos(r \omega_1 t) \Lambda_1 \nonumber\\
&& + E_0 (1-\cos(\omega_1 t)) \Lambda_2. \nonumber
\end{eqnarray}
where the operators $\Lambda_{1,2}$ can be read off as $\Lambda_1 =
\sum_{\ell} n_{\ell}$ and $ \Lambda_{2}= \sum_{\ell} (f_{\ell} +
f_{\ell}^{\dagger})$. Since $|m\rangle$ are eigenstates of
$H_d[J(0), E(0)]$, this allows us to write
\begin{eqnarray}
(i \hbar \partial_t -\epsilon_m) c_m &=&  \sum_{p} c_p \Big[ E_0
(1-\cos(\omega_1 t)) \langle m|\Lambda_2|p\rangle \nonumber\\
&& -J_0 \cos(r \omega_1 t) \langle m| \Lambda_1|p\rangle \Big]
\label{wavexp3}
\end{eqnarray}
These equations are solved numerically for a chain of length $N \le
16$ sites to obtain $c_m(t)$ and hence the wavefunction
$|\psi(t)\rangle$. In our numerics, we find that all results have
similar features for $10 \le N \le  16$. Keeping in mind that
experimental realization of such systems typically involves $N \le
12$, we present our numerical results for $16 \le N \le 12$. In what
follows, we shall use $|\psi(t=T)\rangle$ to compute wavefunction
overlap $F$, the excitation density $n_d$, and the dipole
correlation functions $C_{\ell \ell'}$ which are given by
\begin{eqnarray}
F &=& |\langle \psi(T)|0\rangle|^2 = |c_0(T)|^2 = 1-n_d \label{opexp} \\
C^d_{\ell \ell'} &=& \langle \psi(T)|n_{\ell} n_{\ell'}|\psi(T)
\rangle = \sum_{m p} c_m^{\ast} (T) c_p(T) \langle m| n_{\ell}
n_{\ell'}|p\rangle \nonumber
\end{eqnarray}

\begin{figure}
\includegraphics[height=70mm,width=\linewidth]{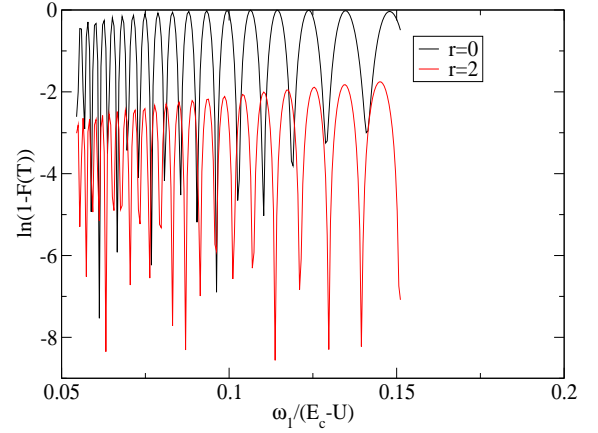}
\caption{Plot of $\ln(1-F)$ as a function of $\omega_1$ for $r=0$
and $2$ for $N=14$. All other parameters are same as in Fig.\
\ref{fig6}.} \label{fig7}
\end{figure}

The results that we obtain from the above mentioned analysis is
shown in Fig.\ \ref{fig6} showing variation of $n_d(T)$ as a
function of $\omega_1$ and $r$ for $N=12$. We find that for most
$r$, $n_d$ is an oscillating function of $\omega_1$ which is
qualitatively similar to its behavior for single rate drive
protocols studied earlier \cite{uma1}. However for $r=2$, $n_d$
displays a minimum for any $\omega_1$; for $r=2$, the value of $n_d$
always ranges between $10^{-2}$ and $10^{-9}$. This behavior can be
clearly seen from Fig.\ \ref{fig9} where $\ln(1-F) = \ln n_d$ is
plotted as a function of $\omega_1$ with $N=14$ and for $r=0$ and
$r=2$. The plot shows that  $\ln n_d \le -2$ for any $\omega_1$ when
$r=2$; thus for $r=2$, the system displays dynamic freezing for a
wide range of $\omega_1$. We find numerically that the overlap
between the final and the initial wavefunctions is closer to unity
for $r=2$ than for any other $r$ and $\omega_1$. We have checked
that such dynamic freezing phenomenon is independent of the system
size $N$ within the range of $N$ covered in our numerical analysis;
$n_d$ shows a dip at $r=2$ for all $N \le 16$.

\begin{figure}
\includegraphics[height=60mm,width=\linewidth]{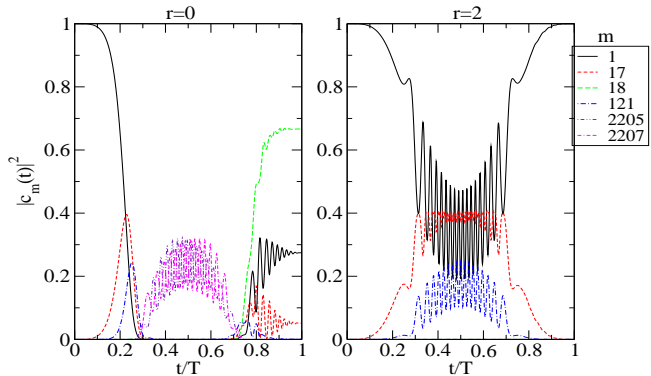}
\caption{Plot of the wavefunction overlap coefficients $|c_m(t)|^2$
as a function of $t$ with $\omega_1/(E_c-U)=0.107$ and $N=16$ for
(a) $r=0$ and (b) $r=2$. We have only plotted coefficients for which
$|c_m(t)|^2 \ge 10^{-4}$. All other parameters are same as in Fig.\
\ref{fig6}.} \label{fig8}
\end{figure}

To understand this phenomenon, we analyze the wavefunction overlap
coefficients $|c_n(t)|^2$ as a function of time $t$ for both $r=0$
and $r=2$. A plot of these coefficients is shown in Fig.\ \ref{fig8}
for $r=0$ and $r=2$. The plot shows that for $r=0$, the system
wavefunction $|\psi(t)\rangle$ has overlap with several eigenstates
$|m\rangle$ (eigenstates of $H(0)$) as shown in left panel of Fig.\
\ref{fig8}. In particular, the change in overlap $|\psi(t)\rangle$
with $|m\rangle$ which occur near $t\simeq T/4, 3T/4$ when the
system traverses the critical point separating the dipole vacuum and
maximal dipole ground states involves several $|m\rangle$ for $r=0$.
Also around these points at $r=0$, $J$ is finite which facilitates
transfer of weight of $|\psi(t)\rangle$ between different
$|m\rangle$ (since these are states with different average dipole
numbers $\langle m| \sum_i n_i |m\rangle$); consequently, one does
not expect the system to return to its initial ground state at $t=T$
for generic $\omega_1$. In contrast, for $r=2$, we find that very
few states ($m=1,\,121,\,{\rm and}\,2207$) contribute to
$|\psi(t)\rangle$. Moreover at $t=T/4, 3T/4$, the transfer of
weights occur only between states $|m=1\rangle$ (dipole vacuum
ground state) and $|m=2207\rangle$ (maximal dipole state for
$N=16$). Thus around these points, the dynamics is controlled by an
effective two-level system similar to those analyzed in Sec.\
\ref{integra}. In addition, we note $J(t)$ vanishes at the critical
point for $r=2$ and remains very small around the critical point;
thus these two levels undergo a nearly unavoided level crossing at
$t= T/4, 3T/4$ since the matrix element of $H(t)$ between states
with different dipole number is extremely small when $J(t) \simeq
0$. This feature, together with the fact only these two states are
involved in dynamics of the system, ensures that $|c_1(t=T)|^2
\simeq 1$ which leads to dynamic freezing. The degree of this
overlap depends on $\omega_1$; however since $J(t)$ always vanishes
at $t=T/4, 3T/4$, the overlap is always close to unity leading to
$\ln(1-F) \le -2$ for the entire range of $\omega_1$ studied in our
numerics.

Finally, we plot the dipole correlation function between sites
$\ell=2$ and $\ell'=0$, $C^d_{20}(T) \equiv C^d_{2}(T)$, as a
function of $r$ and $\omega_1$ in Fig.\ \ref{fig9}. The choice of
the site indices are motivated as follows. Due to the constraint
$n_{\ell} \le 1$ and $n_{\ell} n_{\ell+1}=0$ on the dipole
Hamiltonian, $\langle n_{\ell}^2\rangle = \langle n_{\ell} \rangle$
and $\langle n_{\ell} n_{\ell+1}\rangle=0$ for any $\ell$. Thus the
first non-trivial correlation between dipoles occur for
$|\ell-\ell'|=2$. We find that $C^d_2(T)$ is generically an
oscillating function of $\omega_1$ for any $r$ expect at $r=2$ where
its close to zero for all $\omega_1$. This behavior can be
understood from the fact at $r=2$, $|\psi(t)\rangle$ is close to the
starting ground state $|m=1\rangle$. Since we start our dynamics in
a parameter regime where the ground state is dipole vacuum,
$|m=1\rangle$ and hence $|\psi(t)\rangle$ has no dipole
correlations.

\begin{figure}
\includegraphics[height=60mm,width=\linewidth]{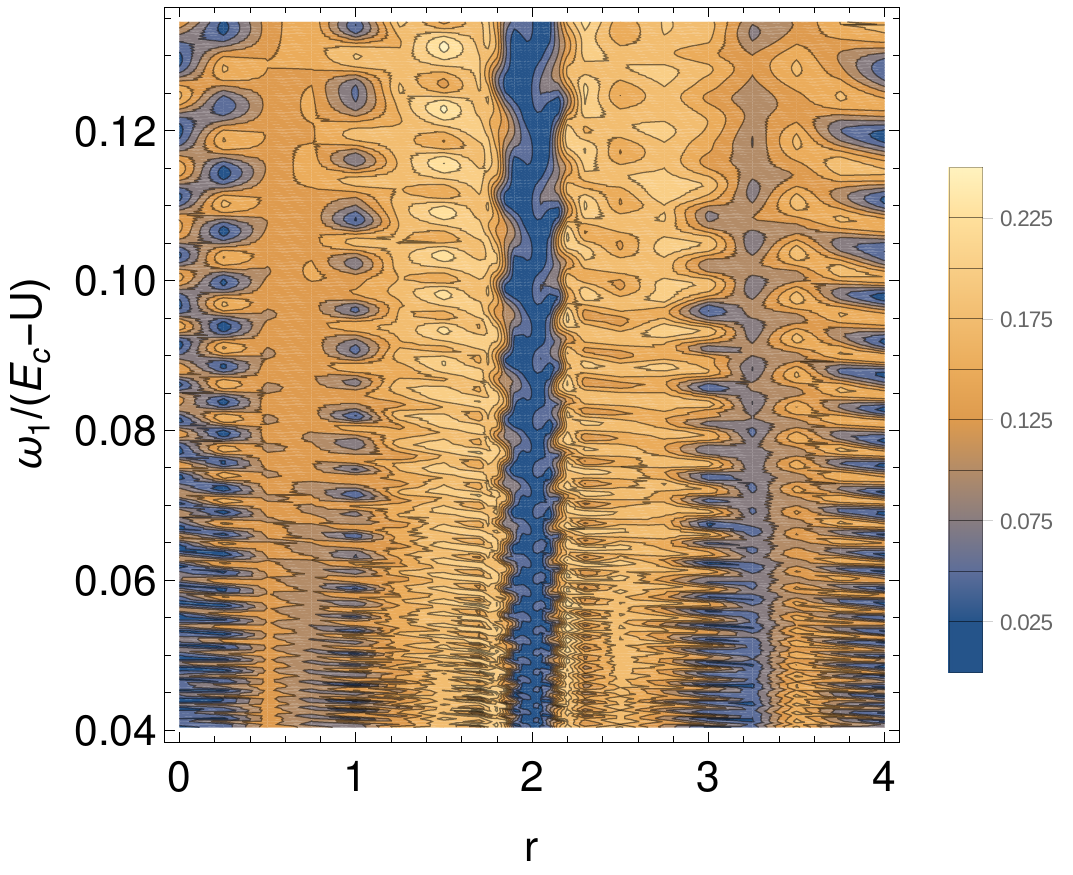}
\caption{Plot of the dipole correlation function $C^d_2(T)$ for
$N=12$ as a function of $\omega_1$ and $r$ showing absence of
correlation at $r=2$ for any $\omega_1$. All parameters are same as
in Fig.\ \ref{fig6}.} \label{fig9}
\end{figure}

\section{Discussion}
\label{diss}

In this work, we have studied periodic dynamics of a class of
integrable [represented by free fermionic Hamiltonian (Eq.\
\ref{fermhamden})] and an experimentally realizable non-integrable
model [represented by bosonic Hamiltonian (Eq.\ \ref{diham1})] using
a two-rate protocol. For numerical purposes, we have chosen the 1D
XY model in a transverse field as a representative example of the
class of integrable model. For each of these models, we find that
the second drive frequency $\omega_2=r \omega_1$ can act as tuning
knob to achieve dynamical freezing.

For integrable models represented by Eq.\ \ref{fermhamden}, we
obtain a semi-analytic expression for the wavefunction at the end of
a drive cycle within the adiabatic-impulse approximation. Our
results constitute a generalization of this method for application
to integrable systems driven by arbitrary two-rate protocols. We
compare the results obtained by this method, for low-enough drive
frequencies where it is expected to be accurate, to exact numerics
and find almost exact match between the two. Our numerics, for these
integrable models, also reveal that the dynamics at $r=1$ is
qualitatively different from those of $r \ne 1$; for $r=1$, dynamic
freezing is obtained for a wide range of $\omega_1$. This behavior
can be analytically explained at a qualitative level. We also show
that the signature of such dynamic freezing can be easily deduced
from fermionic correlators (or equivalently transverse magnetization
of the XY model); the change in magnetization shows a near zero
value at $r=1$ for a wide range of $\omega_1$.

For the tilted Bose-Hubbard model, we drive the applied electric
field and the hopping parameter as a function of time with
frequencies $\omega_1$ and $\omega_2=r \omega_1$ respectively. We
find, similar to integrable models discussed earlier, that $r$ may
act as a tuning parameter for obtaining dynamical freezing; for our
protocol choice, we find that for $r=2$ the system undergoes dynamic
freezing for a wide range of $\omega_1$. We show that for $r=2$, the
wavefunction overlap $F$ satisfies $-2 \ge \ln(1-F) \ge -9$ which
constitutes a much better overlap between the final and the initial
wavefunction than what can be achieved in single-rate protocols
\cite{uma1}. We provide a semi-analytic justification of such
near-perfect dynamic freezing and discuss its signature in
dipole-dipole correlation function.

Experimental verification of our result could be most easily
achieved by on-site parity of boson occupation ($(-1)^{n_i}$)
measurement of bosons in the tilted Bose Hubbard model after a drive
cycle. Such measurement have already been carried out for tilted
bosons in equilibrium for $N=12$ sites \cite{bakr1}. Here we propose
to start from the dipole vacuum $(n_d=0$) state which correspond to
odd ($n=1$) boson occupation on all sites. We then propose to drive
the system for one full period using the two-rate protocol given by
Eq.\ \ref{prot1} and measure the number of sites with even bosons.
Since formation of a dipole in this case always lead to even
($n=0,2$) boson occupation at adjacent sites, such a measurement
provide us with information on dipole density and hence $1-F$. We
predict that for $r=2$, almost no dipoles would be created for a
wide range of $\omega_1$.

In conclusion, we have studied two-rate periodic protocols applied
to a class of integrable models and an experimentally realizable
non-integrable models. Our study reveals that the second drive
frequency may be used as a tuning parameter to achieve dynamical
freezing. We have provided a detailed analysis of this phenomenon,
analyzed their signature in correlation functions, and suggested
concrete experiments to test our theory.

{\it Acknowledgement}: KS thanks J.D. Sau for discussion on related
projects. SK acknowledges financial support from CSIR, India, under
Scientists' Pool Scheme No. 13(8764-A)/2015-Pool.

\end{document}